\documentclass[doublespacing]{elsart}
\usepackage{graphicx}
\usepackage{bm} 
\usepackage{amssymb}
\usepackage{lineno}
\usepackage[ps2pdf,colorlinks=true]{hyperref}
\hypersetup{backref,colorlinks=true}
\journal{Nucl. Instr. and Meth. in Phys. Res. A}
\begin{document}
\begin{frontmatter}
\title {A large area plastic scintillator detector array for fast neutron measurements}
\author[a,b]{P. C. Rout },
\author[a,b]{D. R. Chakrabarty},
\author[a,b]{V. M. Datar},
\author[a]{Suresh Kumar},
\author[a]{E. T. Mirgule},
\author[a]{A. Mitra},
\author[c]{V. Nanal} and
\author[a]{R. Kujur}
\address[a]{Nuclear Physics Division, Bhabha Atomic Research Centre, Mumbai-400 085, India}
\address[b]{Homi Bhabha National Institute, Bhabha Atomic Research Centre, Mumbai-400 085, India}
\address[c]{Department of Nuclear and Atomic Physics, Tata Institute of Fundamental Research, Mumbai-400 005, India}

\begin{abstract}
A large area plastic scintillator  detector array($\sim$ 1$\times$1
m$^2$) has been set up for fast neutron spectroscopy at the
BARC-TIFR Pelletron laboratory, Mumbai. The energy, time and
position response has been measured for electrons using radioactive
sources and for mono-energetic neutrons using the
$^7$Li(p,n$_1$)$^7$Be*(0.429 MeV) reaction at proton energies
between 6.3 and 19 MeV. A Monte Carlo simulation of the energy
dependent efficiency of the array for neutron detection is in
agreement with the $^7$Li(p,n$_1$) measurements. The array has been
used to measure the neutron spectrum, in the energy range of 4-12
MeV, in the reaction $^{12}$C+ $^{93}$Nb at E($^{12}$C)= 40 MeV.
This is in reasonable agreement with a statistical model
calculation.

\end{abstract}
\end{frontmatter}
\section{Introduction}

The measurement of neutron spectra is important in many nuclear
reaction studies. One example is the study of the statistical decay
of compound nuclei populated in low energy fusion reactions. In
particular, it would be interesting to make a measurement in the
$^{12}$C+$^{93}$Nb reaction because an unusual structure has been
reported in the angular momentum gated proton and alpha spectra in
this reaction at E($^{12}$C) = 40 and 42.5 MeV \cite{mitra1,mitra2}.
This would require measurements of neutron spectra down to
$\sim$~0.1~$\mu$b~sr$^{-1}$~MeV$^{-1}$ with an energy resolution
$\triangle$E$_{FWHM}$/E $\lesssim$~10$\%$. A large area neutron
detector is needed to meet these requirements. Another example of
interest is the study of shell effects, and their damping with
excitation energy, on the nuclear level density(NLD) in the region
of doubly closed shell nuclei. The variation of NLD for a wide range
of the N/Z ratio, where N and Z are the neutron and proton numbers
in a nucleus, using radioactive ion beams is an interesting topic
that can be addressed. A large neutron detector array would also be
useful in coincidence measurements involving neutrons, e.g. study of
fission dynamics and neutron decay following the transfer reaction.

Plastic scintillators have been widely used for neutron measurements
by the time of flight (TOF) technique because of their fast response
and the relatively low cost, enabling the construction of a large
detector system. A long scintillator detector with photomultiplier
tubes (PMTs) at either end was first used by Charpak et
al.\cite{charp}. While economizing on the number of PMTs for a given
areal coverage, the detector was shown to have good TOF and position
resolution using the timing information from both the PMTs. An array
consisting of a number of such long scintillators stacked one above
the other is a cheaper and simpler alternative to an array of a
larger number of discrete detectors with a comparable efficiency.
For example a 1~m$^2$ array with 16 bars of cross section 6~cm
$\times$~6~cm needs 32~PMTs of 5~cm diameter. An array of similar
overall dimensions but more granularity could have 256 plastic
scintillators each of size 6~cm$\times$6~cm and would require 256
PMTs. An array of 49 square plastic scintillators of size
14~cm$\times$14~cm could be viewed by 12.7~cm diameter PMTs or 5~cm
diameter PMTs with light guides. Either option would increase the
complexity and cost of the detector system. We have chosen to set up
a 1~m$\times$1~m plastic detector array consisting of 16 long
scintillators.

The paper is organized in four sections. The first section gives
details of the detector array. In the next section the measured
time, position and energy response to electrons and monoenergetic
neutrons is described. The electron measurements were carried out
using radioactive sources while those for neutrons were done using
the $^7$Li(p,n$_1$)$^7$Be*~(0.429 MeV) reaction at proton energies
from 6.3 to 19~MeV. The third section outlines the Monte Carlo
simulation of the response of the plastic detector to fast neutrons.
Finally, an example of a measured neutron spectrum in the reaction
$^{12}$C+$^{93}$Nb at E($^{12}$C)~=~40~MeV is presented to
demonstrate the capability of the neutron detector array.
%
\begin{figure}[!h]
\begin{center}
\includegraphics[scale=0.5]{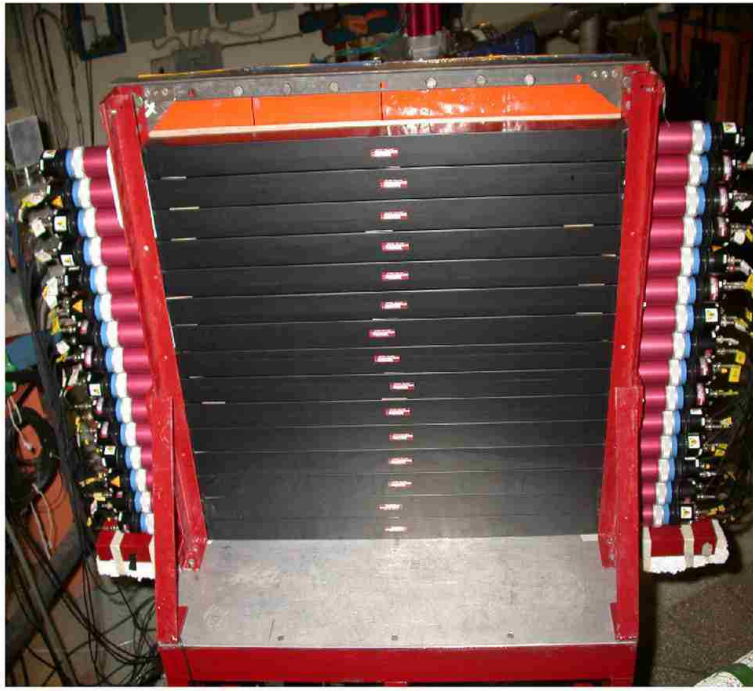}
\caption{A photograph  of the plastic scintillator detector array
taken with 15 bars.}
\label{fig:fig01}
\end{center}
\end{figure}

\section{Description of Neutron Detector Array}

The neutron detector array consists of 16 plastic scintillator bars
(equivalent to Bicron BC-408 and procured from SCIONIX, Holland) of
square cross section\cite{scionix}. Each bar has a dimension
6~cm$\times$6~cm$\times$100~cm and is coupled to two 5~cm diameter
XP2020 PMTs, one each at either end. The scintillator has a light
output of $\sim$~65\% compared to that of anthracene, a
scintillation decay time of $\sim$4~ns and a bulk attenuation length
$>$~3~m. The polyvinyl toluene based plastic scintillator has a
carbon to hydrogen ratio of $\sim$1:1.1. The density and refractive
index of the scintillator are 1.03~gm/cm$^3$ and 1.58, respectively.
The spectral sensitivity of the XP2020 PMT peaks at 420 nm, with a
quantum efficiency of $\sim$ 25~\%, and matches the emission
spectrum of the plastic scintillator. The PMTs have a fast response
time (rise time $\sim$1.3~ns) and a gain of $\sim$10$^7$ at about
2~kV bias voltage. The PMTs are powered by a 32 channel programmable
high voltage power supply developed in-house~\cite{manna}. The
plastic scintillators are stacked one above the other on a stand
with wheels for horizontal movement of the array and a four bolt
arrangement to adjust the height and level. A photograph of the
array is shown in Fig.~\ref{fig:fig01}.

Lead (Pb) sheets of total thickness 25~mm can be placed in front of
the array to reduce the low energy $\gamma$-ray background while not
significantly attenuating the neutrons from the target. Apart from
the target neutrons directly reaching the detector, there are those
following a circuitous path, such as scattering from various
materials in the experimental hall. In order to compare the
contributions from these two sources, a 30~cm thick mild steel (MS)
shield consisting of several plates of increasing transverse
dimension was fabricated. This can be placed near the target to
block the target neutrons from reaching the array. The measurements
with and without the MS shield allows an estimation of the
contribution due to the scattered neutrons that are not in the line
of sight of the target.
\begin{figure}[!t]
\begin{center}
\includegraphics[scale=0.80]{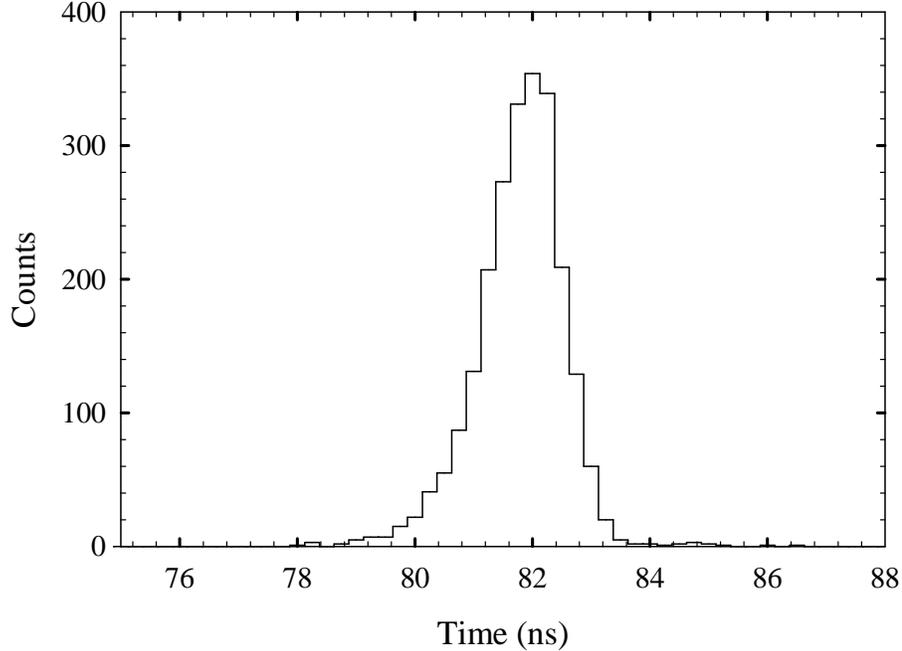}
\caption{TOF spectrum,with an arbitrary offset(see eq.1)  derived
from the left and right time (T$_L$,T$_R$) of PMTs by exposing the
centre of the detector to 511 keV photons from a $^{22}$Na source.}
\label{fig:fig02}
\end{center}
\end{figure}
\section{Time, position and energy response}

 The important parameters deduced from the time and integrated charge
 of the PMT signals are as follows\\
%
\begin{figure}[!t]
\begin{center}
\includegraphics[scale=0.85]{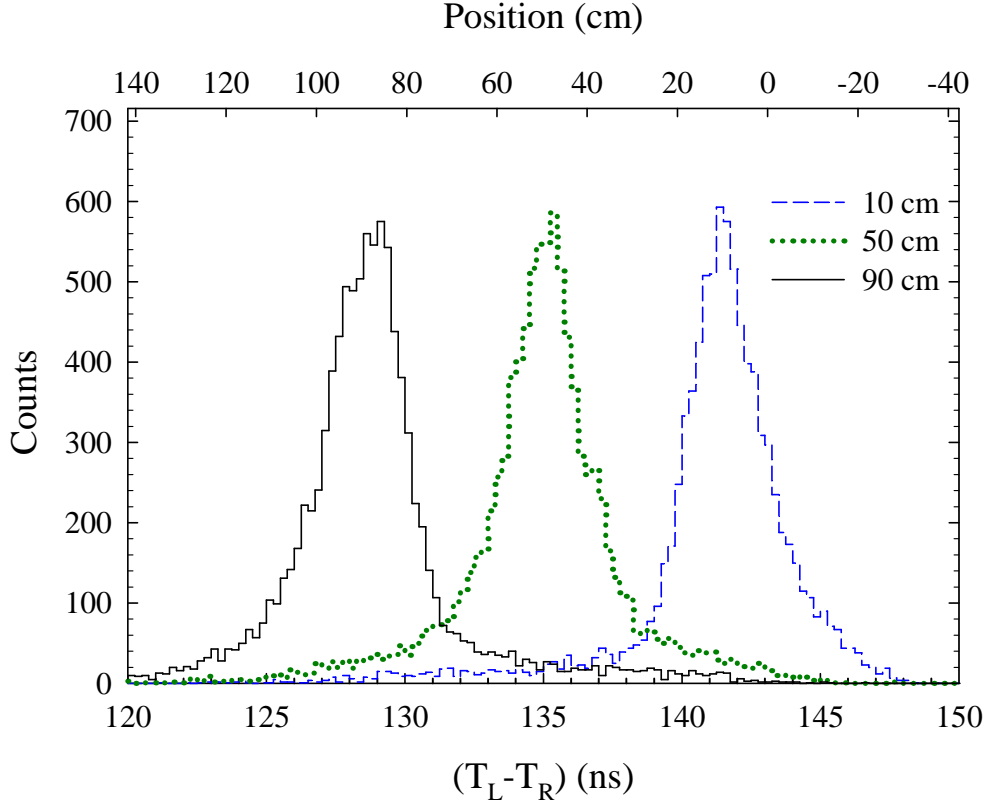}
\caption{ Position response measured by exposing $\sim$~1~cm wide
portion along the length of the scintillator bar, using lead
collimators and a $^{22}$Na $\gamma$-ray source, for various
distances of the exposed part from one end. The upper axis shows the
distance calculated using the fit shown in Fig.~\ref{fig:fig04}. The
lower axis shows the measured time difference (T$_L -$~T$_R$).}
\label{fig:fig03}
\end{center}
\end{figure}

(a) The time of flight(TOF), derived from left and right PMT trigger
times (T$_L$, T$_R$), with respect to a reference time, is given by
\begin{equation}
TOF = (T_L+T_R)/2 + T_{offset}
\end{equation}
where T$_{offset}$ is independent of the position of the interaction
point.
%
\begin{figure}[!t]
\begin{center}
\includegraphics[scale=0.85]{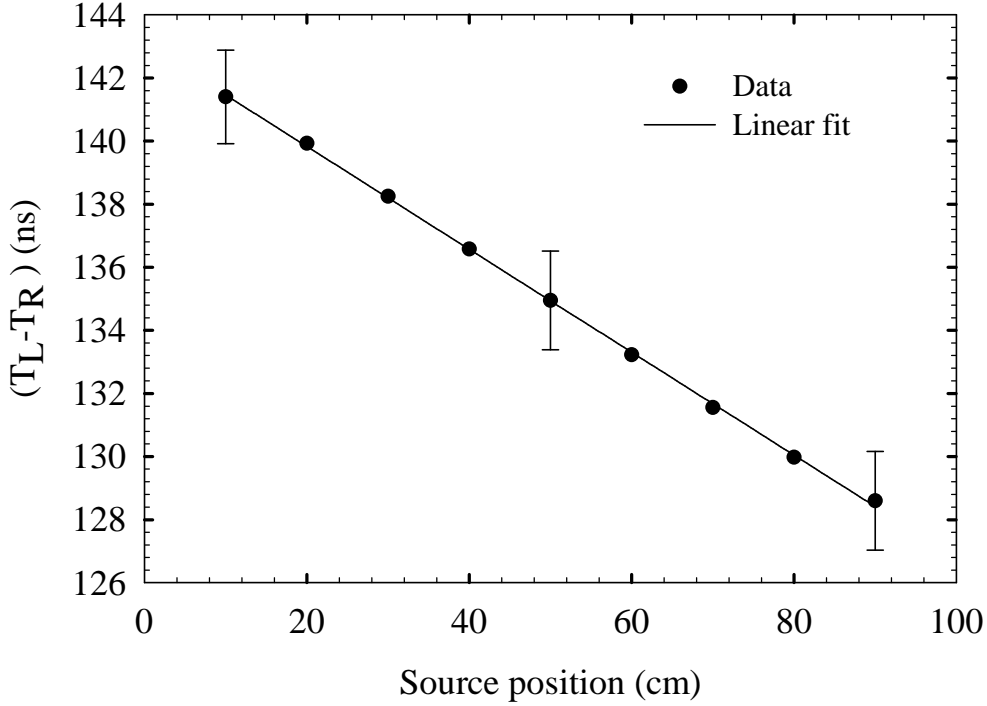}
\caption{ Measured (T$_L -$~T$_R$) for various source positions
along the length of the scintillator. The vertical lines (shown only
for three points) represent FWHM of the position response shown in
Fig.~\ref{fig:fig03}.} \label{fig:fig04}
\end{center}
\end{figure}

(b) The position (X) of interaction point is obtained from the time
difference between T$_L$ and T$_R$
\begin{equation}
X \propto (T_L-T_R).
\end{equation}
In addition, the position information can also be obtained from the
integrated charges of the left and right PMT signals (Q$_L$, Q$_R$)
as
\begin{equation}
X \propto ln \,\frac{Q_L}{Q_R}.
\end{equation}
%
\begin{figure}[!t]
\begin{center}
\includegraphics[scale=0.85]{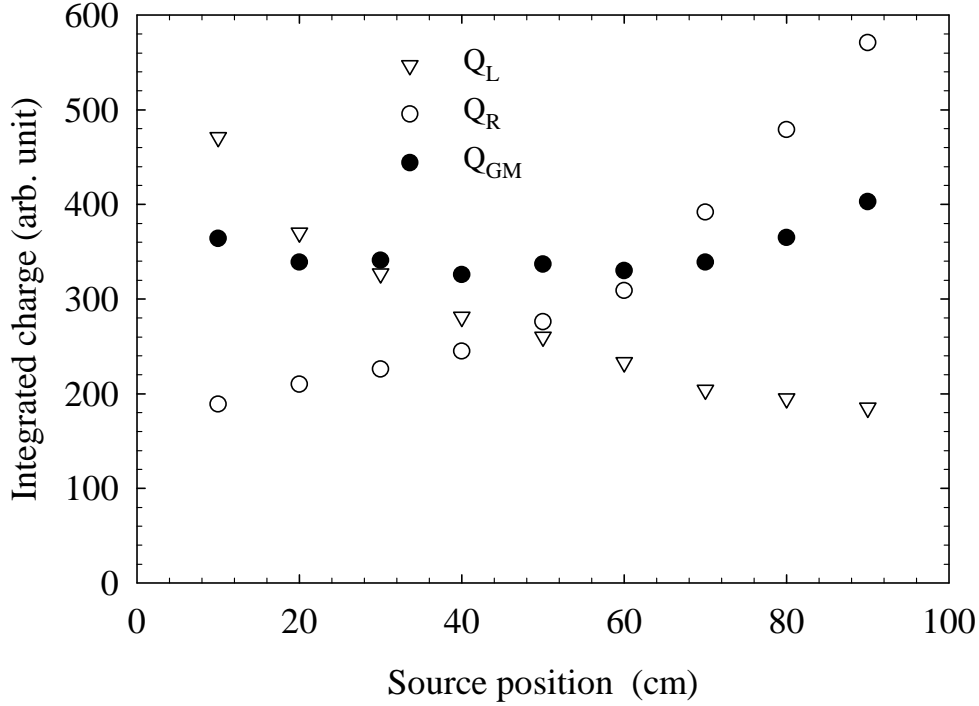}
\caption{Variation of $Q_L, Q_R$ and $Q_{GM}$ (see text) as a
function of distance from one end.}
\label{fig:phtvar}
\end{center}
\end{figure}

(c) The integrated charge information is necessary for the
determination of neutron detection efficiency. The geometrical mean
$Q_{GM}$ of two signals
\begin{equation}
Q_{GM}=\sqrt{Q_L\,Q_R}
\end{equation}
is roughly independent of the position and is proportional to the
energy deposited in the detector (see for example
Ref.\cite{karsch}).
%
\begin{figure}[!t]
\begin{center}
\includegraphics[scale=0.9]{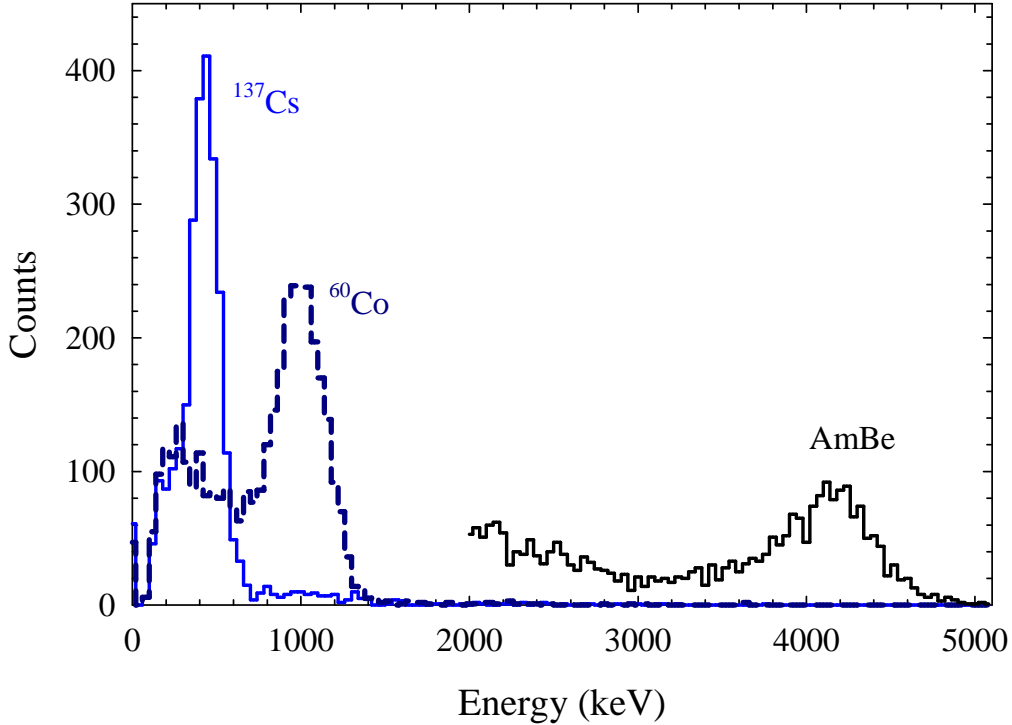}
\caption{Energy spectra measured in the plastic scintillator for
$\sim$450~keV, 1~MeV and 4.15 MeV  electrons using various
radioactive sources illuminating the central position.}
\label{fig:eresp-e}
\end{center}
\end{figure}
%
%
\begin{figure}[!t]
\begin{center}
\includegraphics[scale=0.9]{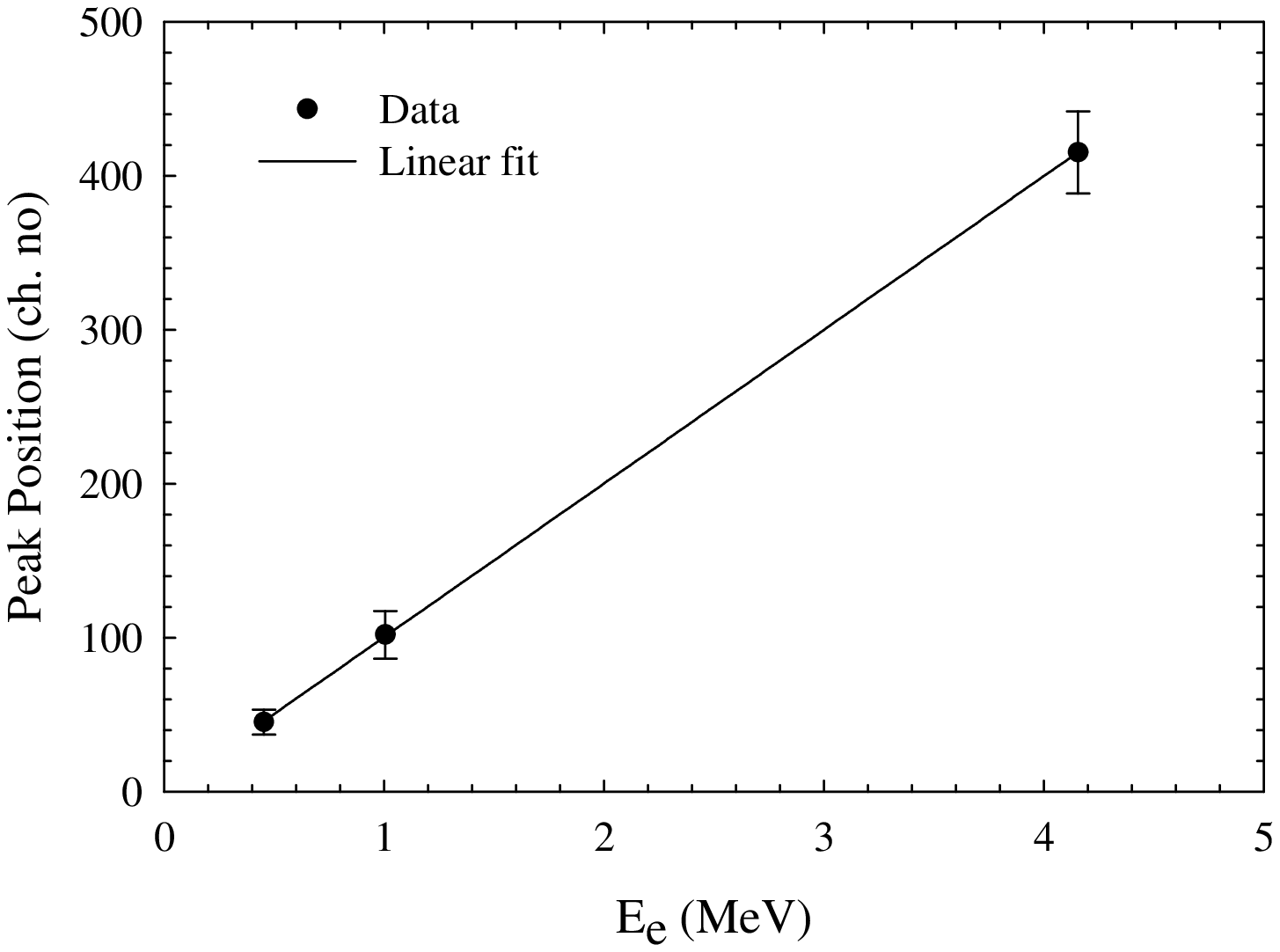}
\caption{Electron energy response of the plastic scintillator. The
vertical error bars depict the FWHM of the peaks shown in
Fig.~\ref{fig:eresp-e} } \label{fig:elinearity}
\end{center}
\end{figure}
%

\subsection {Response to electrons}

The time, position and energy response of the plastic scintillator
was measured with collimated $\gamma$-rays from various radioactive
sources.
The 511 keV $\gamma$-ray from $^{22}$Na was detected in a plastic
bar in coincidence with the complementary 511 keV $\gamma$-ray
detected in a 8~cm thick BaF$_2$ detector with a hexagonal cross
section and having a face to face distance of 6~cm. Both T$_L$ and
T$_R$, measured with respect to BaF$_2$, were recorded. The TOF and
position information were extracted using the expressions given
earlier. The TOF spectrum obtained is shown in Fig.~\ref{fig:fig02}.
The full width at half maximum (FWHM) is $\sim$~1.4~ns, which
includes the contribution from the electronics as well as the
transit time spread of PMTs. Fig.~\ref{fig:fig03} shows the position
spectrum derived from T$_L$ and T$_R$. The position resolution is
$\sim$~20~cm and the response is linear as shown in
Fig.~\ref{fig:fig04}. For a given amount of energy deposition the
integrated charge from a PMT varies by a factor of $\sim$3 from one
end to other as shown in Fig.~\ref{fig:phtvar}. The geometric mean
$Q_{GM}$ is almost independent of position, within 5\%, except at
the ends where it increases up to $\sim$~15\%.

The energy resolution for mono-energetic electrons produced within
the plastic scintillator was measured using the response to a
recoiling electron in the Compton scattering of $\gamma$-rays. The
back scattered photon was detected in a BaF$_2$ detector for tagging
on the energy of the electron. $\gamma$-rays from radioactive
sources $^{137}$Cs, $^{60}$Co and $^{241}$Am-Be were collimated
using appropriately placed lead bricks to illuminate a 2~cm wide
portion of the plastic scintillator.
%
\begin{figure}[!t]
\begin{center}
\includegraphics[scale=0.95]{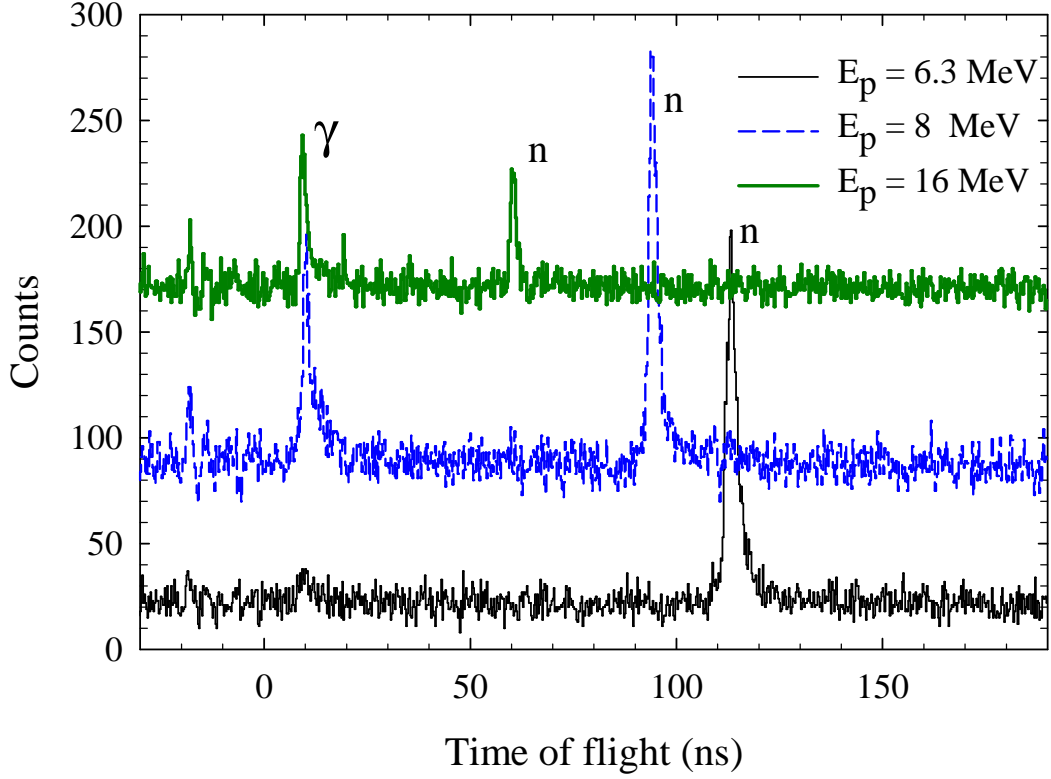}
\caption{TOF spectra from p~+~$^{nat}$Li reaction at E$_p$~=~6.3, 8
and 16 MeV. The position of the gamma and neutron peaks are
indicated.}
\label{fig:tofn1}
\end{center}
\end{figure}
The source and collimator assembly was moved along the length of the
plastic to measure the position dependence of the response. The
measured energy spectra of the electrons for the central source
position are shown in Fig.~\ref{fig:eresp-e}. The energy response is
linear as shown in Fig.~\ref{fig:elinearity}. The
$\triangle$E$_{FWHM}$/E was measured to be $\sim$~35\% and $\sim$~12\%
for 450~keV and 4.15~MeV electrons, respectively, produced at the
centre of the plastic scintillator. This was almost independent of
the position in the plastic. The finite size of the BaF$_2$ detector
and the illuminated zone in the plastic had an insignificant
contribution to the measured energy resolution.
%
\begin{figure}[!t]
\begin{center}
\includegraphics[scale=0.95]{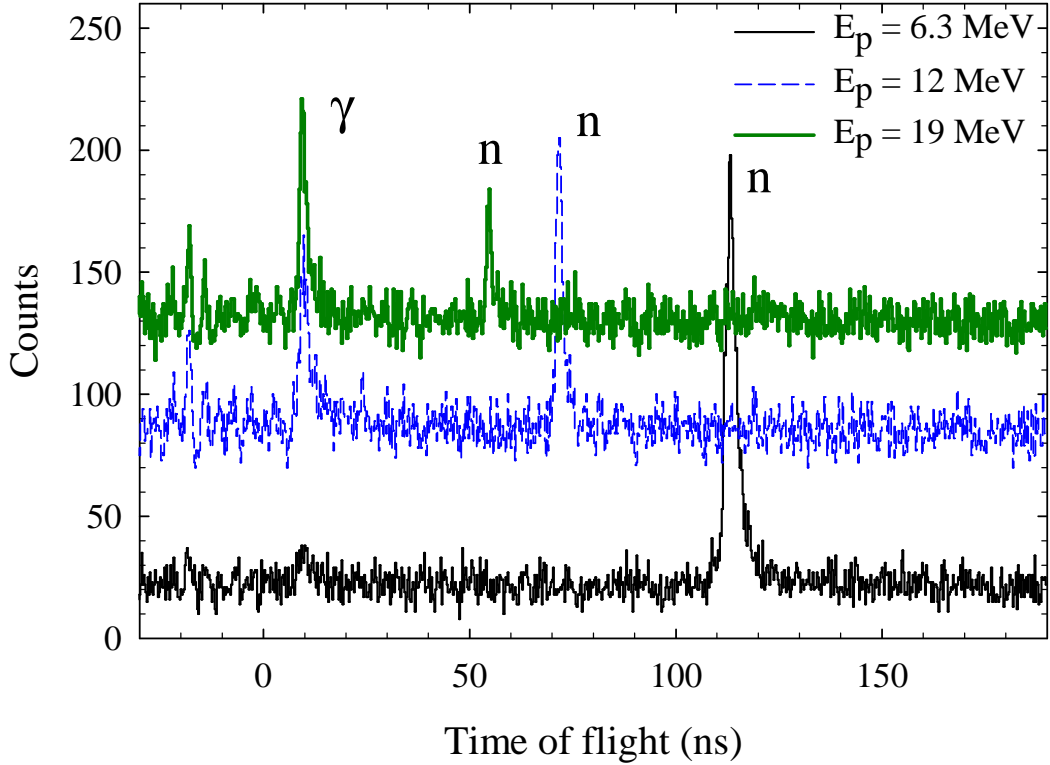}
\caption{ Same as in Fig.~\ref{fig:tofn1} for E$_p$~=~6.3, 12.0 and
19 MeV. }
\label{fig:tofn2}
\end{center}
\end{figure}
%
\subsection {Response to monoenergetic neutrons}
Monoenergetic neutrons were produced by bombarding a
$\sim$~2~mg/cm$^2$ thick $^{nat}$Li metal target with proton beams
of energies 6.3, 8, 12, 16 and 19 MeV from the Mumbai Pelletron. The
neutrons produced in the reaction $^7$Li(p,n$_1$)$^7$Be$^\ast$ were
detected by one of the plastic scintillators in coincidence with the
429 keV $\gamma$-ray emitted from the first excited state of $^7$Be.
The $\gamma$-ray was measured in the BaF$_2$ detector, mentioned
earlier, placed at $\sim$~2~cm from the target. The plastic
scintillator detector was placed at a distance of 3~m from the
target and at 45$^\circ$ with respect to the beam direction. The TOF
of the events in the plastic scintillator was measured with respect
to the BaF$_2$ detector using time to amplitude converters~(TAC)
calibrated using a high precision time calibrator.
%
\begin{figure}[!t]
\begin{center}
\includegraphics[scale=0.9]{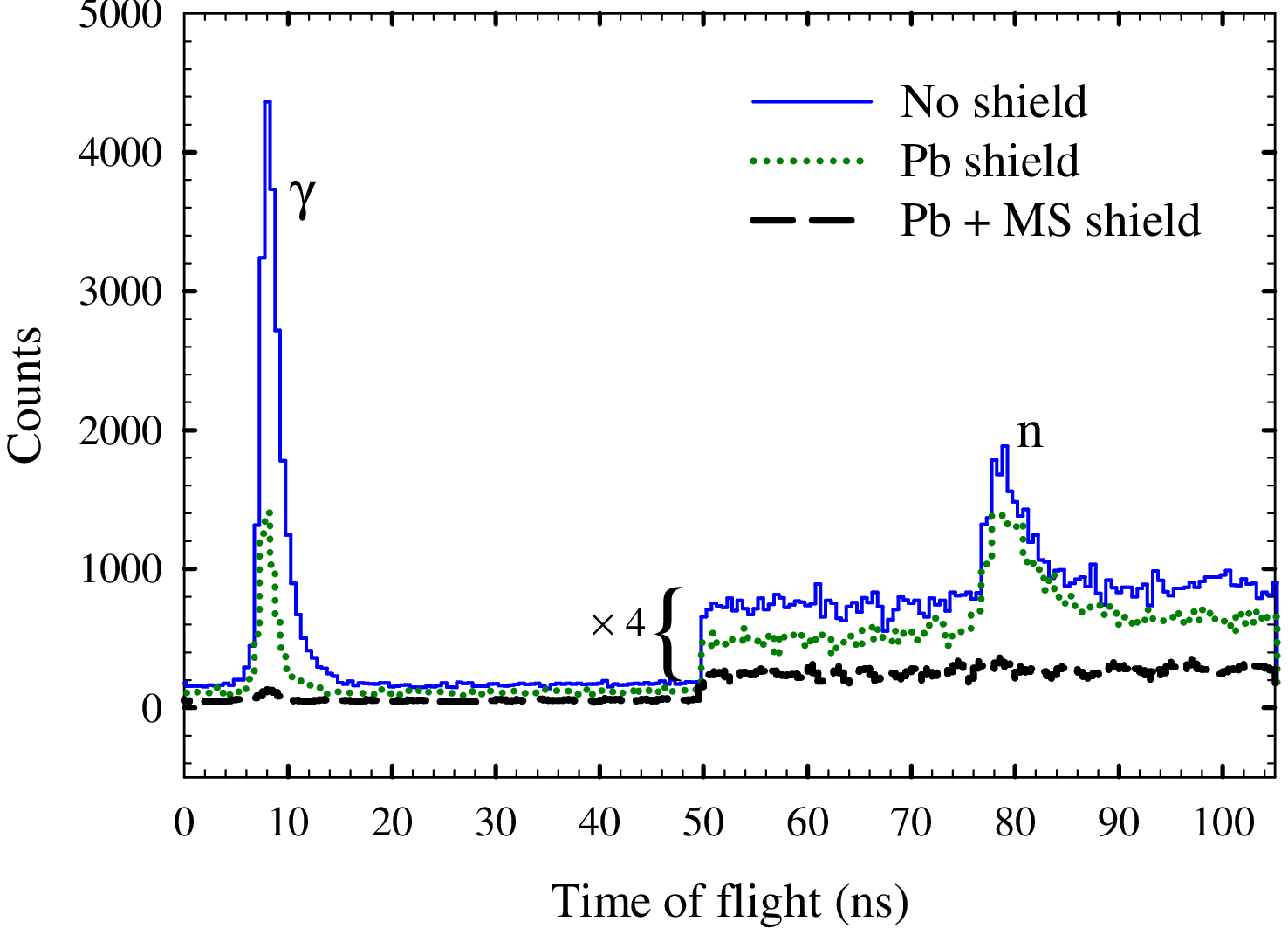}
\caption{TOF spectra with and without lead and MS shields for a
plastic scintillator.}
\label{fig:gam-n-shield}
\end{center}
\end{figure}
The measured TOF spectra for the neutron group populating directly
the 0.429 MeV excited state in $^7$Be are shown in
Figs.~\ref{fig:tofn1} and \ref{fig:tofn2}. The peaks in the TOF
spectra correspond to neutrons of energies 3.7, 5.3, 9.0, 12.7, 15.4
MeV, respectively. The relative time differences between the
neutrons and $\gamma$-ray peaks decrease with increase in neutron
energy and agree with the expected values within 2\%.

In order to assess the attenuation of neutron and $\gamma$-rays by
the lead shield a measurement was made  at E$_p$=6.3~MeV. In the
same measurement the contribution of scattered neutrons were
assessed by placing the MS shield near the target. The TOF spectra
are shown in Fig.~\ref{fig:gam-n-shield}. A comparison of the
spectra shows that the lead shield attenuates the neutron group by
$\sim$~10\% and the $\gamma$-rays by $\sim$70\%. Moreover, the
contribution from the scattered neutrons is negligible (with an
upper limit of 5\%).
\begin{figure}[!t]
\begin{center}
\includegraphics[scale=0.9]{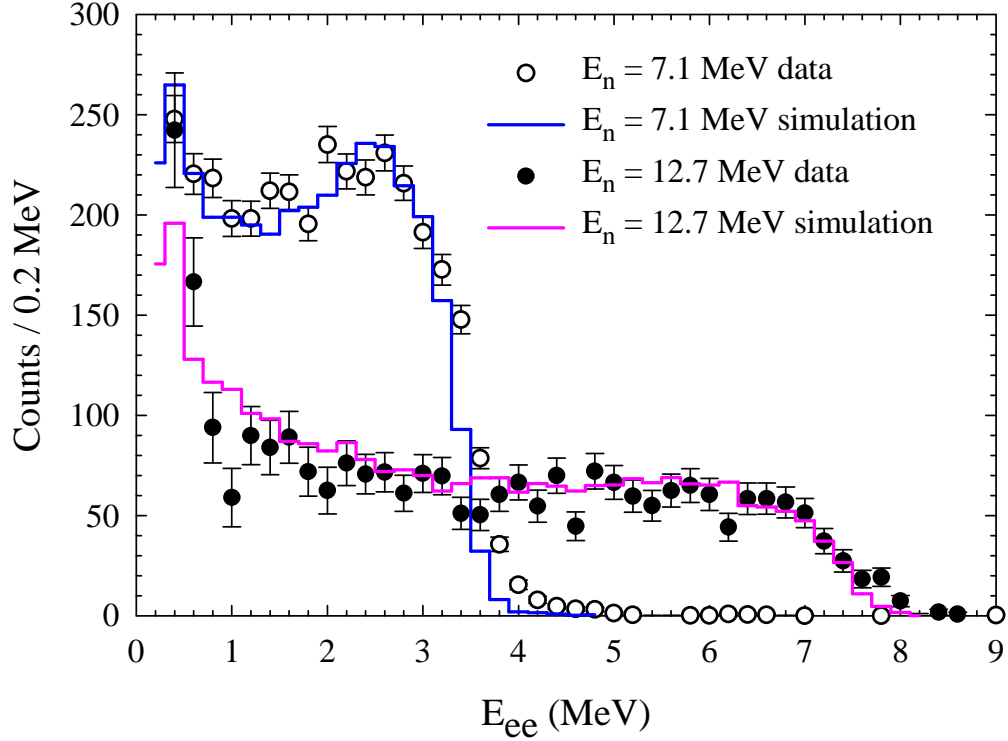}
\caption{Experimental and simulated energy (electron equivalent)
response of the plastic scintillator to neutrons of energies 7.1 and
12.7 MeV.} \label{fig:eresp-n}
\end{center}
\end{figure}
\subsection {Efficiency measurement for monoenergetic neutrons}

The efficiency  of plastic detector for monoenergetic neutrons  was
measured using the same reaction. The proton beam energies were 10
and 16~MeV  corresponding to neutron energies of 7.1 and 12.7~MeV,
respectively. One plastic bar was used to measure neutrons in
coincidence with 429 keV $\gamma$-rays detected in an array of seven
close-packed hexagonal BaF$_2$ detectors of dimensions mentioned
earlier. The plastic detector was placed at 45$^\circ$ with respect
to the beam and 1.5~m from the target. The BaF$_2$ array was placed
$\sim$ 2 cm from the target. The anode signal from PMT of each
BaF$_2$ detector was amplified and split to measure energy and
pileup. The split signals were fed to two charge sensitive analog to
digital converters (QDC) with gate widths of 2~$\mu$s and 200~ns.
The dynode signals from both left and right PMTs of the plastic
scintillator were amplified for pulse height response and the anode
signals were processed to measure TOF with respect to the logical OR
of BaF$_2$ timing signals. The energy calibrations of plastic and
BaF$_2$ detectors were done using $^{137}$Cs and $^{60}$Co
$\gamma$-ray sources. The energy spectrum of the plastic detector
for a given neutron energy was obtained by gating on the
corresponding peak in the TOF spectrum and subtracting the random
contribution. These are shown in Fig.~\ref{fig:eresp-n}. Here the
energy corresponds to the electron equivalent energy~($E_{ee}$) of
the charged particles produced by neutron interaction in the
scintillator.

The efficiencies of the plastic scintillator for the two neutron
energy groups mentioned above were estimated as follows. In the data
analysis the number~(M) of BaF$_2$ detectors detecting the 429~keV
$\gamma$-ray in coincidence was generated event by event. The events
with M$>$1 obviously correspond to random events. The yield of the
n$_1$ group at each proton energy was obtained from the TOF spectra
for the condition of M=1 and E$_{ee}\ge$250~keV. The neutron
detection efficiencies were obtained from these yields, the target
thickness and the angle dependent cross section of the n$_1$ group.
This cross section was taken from Poppe {\it et al.}~ \cite{pope}.
The target thickness was obtained from the singles $\gamma$-yield in
all seven BaF$_2$ detectors measured at 16 MeV proton energy. The
efficiency of the BaF$_2$ array for 429~keV $\gamma$-rays was
calculated as 26.7 \% using the Electron Gamma Shower (EGS)
simulation program ~\cite{nelson}. The target thickness extracted
using the angle integrated cross section from Ref.\cite{pope} was
2.8 $\pm$ 0.3~mg/cm$^2$. The neutron detection efficiencies
estimated from the above procedure were divided by the geometrical
efficiency of the detector to get the intrinsic efficiencies. These
are shown in Table~\ref{T1}. These measured efficiencies, however,
include the contributions of the scattered neutrons from the BaF$_2$
array. This effect was estimated to be $<$~7\% from the Monte Carlo
simulation discussed in the next section.

\begin{table}
\caption{ Measured and simulated intrinsic efficiency of plastic
scintillator detector for mono-energetic neutrons in the absence~(A)
and presence~(B) of the BaF$_2$ array.}
\label{T1}
\begin{center}
\begin{tabular}{|c|c|c|c|}
\hline E$_n$(MeV) & Measured  & Simulated & Simulated \\
&efficiency (\%)& efficiency(\%)~(A) &efficiency(\%)~(B)\\
\hline
7.1 & 27.4 $\pm$ 2.8 & 26.4 &27.2\\
\hline
12.7 & 22.7 $\pm$ 3.2 &20.8 & 22.3\\
\hline
\end{tabular}
\end{center}
\end{table}

\section{Monte Carlo simulation of the neutron detector array}

The simulation of the response of the plastic detector array  to
fast neutrons has been performed by a Monte Carlo program developed
in our laboratory. The processes by which neutrons of energy up to
$\sim$~20~MeV interact with the detector material were taken as (a)
elastic scattering on hydrogen and $^{12}$C, (b) inelastic
scattering to the 4.4 and 7.6 MeV states in $^{12}$C and (c)
$^{12}$C(n,$\alpha$), $^{12}$C(n,p) reactions. Among these the first
process is the dominant one. The cross sections were taken from
Ref.~\cite{cecil}.

In the program, neutrons of a given energy are made incident on the
plastic array, placed at a certain distance from the source,
assuming angular isotropy. The energy dependent total interaction
cross section determines whether the neutron scores a hit in the
detector. If it scores a hit, the interaction process is chosen
randomly on the basis of the relative magnitude of the respective
cross section. If the process is elastic scattering on hydrogen, the
scattering angle in the centre of mass (c.m.) system is chosen
randomly. This decides the energy of the recoiling proton which is
assumed to be fully deposited in the detector. In (n,p),
(n,$\alpha$) and (n,n$^\prime)3\alpha$ processes, the energy of p
and $\alpha$ were calculated assuming isotropic emission in the c.m.
system. For the (n,n$^\prime\,\gamma)$ process, the 4.4~MeV
$\gamma$-ray deposits energy depending on the attenuation length and
the available path in the scintillator for a given propagation
direction. The electron energy equivalent~(E$_{ee}$) of the
deposited energy for heavy charged particles are obtained using the
following empirical expression
\begin{equation}
T_e=a_1T_p-a_2\big[1.0-exp(-a_3{T_p}^{a_4})\big].
\end{equation}
The choice of the parameters a$_1$ to a$_4$ was guided by those
given by Cecil {\it et al.}~\cite{cecil} with some fine tuning.
These are presented in Table~\ref{T2}.
\begin{table}
\caption{Parameters in the light response function for proton and
alpha.}
\label{T2}
\begin{center}
\begin{tabular}{|c|c|c|c|c|}
\hline
Particles & a$_1$  & a$_2$ & a$_3$ & a$_4$ \\
\hline
Proton  & 0.97 & 7.6 & 0.1 & 0.90 \\
\hline
Alpha & 0.41 & 5.9 &0.065 & 1.01\\
\hline
\end{tabular}
\end{center}
\end{table}
The light output due to recoiling $^{12}$C nuclei is small and
therefore neglected. The time of occurrence of the PMT signal was
derived from the neutron flight times and the scintillation photon
propagation time. The simulated left and right PMT times (T$_L$,
T$_R$) and E$_{ee}$  from each event were  used to generate TOF and
E$_{ee}$ spectra provided E$_{ee}$ is greater than the experimental
threshold. The TOF spectrum for a given neutron energy shows a peak
with a width arising from the intrinsic time resolution of the
plastic scintillator as well as the variation in the interaction
position. The shapes of the simulated E$_{ee}$ spectra at neutron
energies of 7.1 and 12.7 MeV compare well with the experiment as
shown in Fig.~\ref{fig:eresp-n}.

\begin{figure}[!t]
\begin{center}
\includegraphics[scale=0.85]{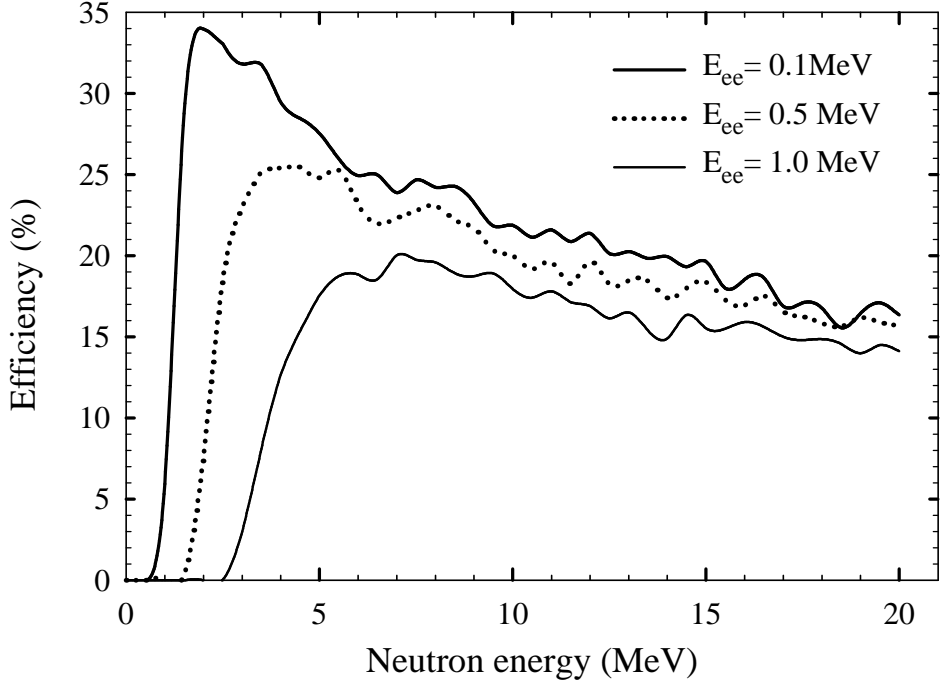}
\caption{Simulated neutron efficiency of a plastic bar for different
energy thresholds E$_{ee}$.}
\label{fig:neff-sim}
\end{center}
\end{figure}

The ratio of detected events to the number of incident neutrons
gives the intrinsic neutron efficiency. The simulated intrinsic
efficiencies of one plastic scintillator for neutrons of energies
7.1 and 12.7~MeV  were obtained for the experimental
threshold~(E$_{ee}$) of 250 keV. These are shown in the third column
of Table~\ref{T1}. The energy dependent neutron efficiencies for
various thresholds are shown in Fig.~\ref{fig:neff-sim} for a single
plastic scintillator with the lead shield in front. It can be seen
that while an increase in E$_{ee}$ does not substantially change the
neutron detection efficiency at higher energies, it increases the
effective threshold energy for neutron detection.

The presence of the BaF$_2$ array near the target is a source of
scattered neutrons, some of which reach the plastic detector. In
order to estimate their contribution to the measured efficiencies,
the simulation program was extended to include the elastic and
non-elastic processes from Ba and F in BaF$_2$. The increase in the
TOF of the elastically scattered neutrons is too small to
distinguish from that of the neutrons directly reaching the plastic.
The non-elastic processes lead either to a loss of neutrons as in
the (n,p), (n,$\alpha$), (n,$\gamma$) reactions or to a low energy
neutrons from the (n,n$^\prime$) and (n,2n) reactions. These
processes, therefore, do not contribute to the counts in the TOF
peak. The intrinsic efficiencies in the presence of BaF$_2$ are
renormalised by the extra contribution in the TOF peak essentially
due to elastic scattering. These are shown in the last column of
Table~\ref{T1}. The renormalised efficiencies are increased by
$<$~7\%. These are however the upper limits because we have assumed
isotropy in the scattering processes whereas the limited elastic
scattering data shows forward peaked distribution (see for example
 Ref.~\cite{kuijper}). It may be mentioned that the elastic
 scattering process on BaF$_2$ does not lead to any significant change in
 energy spectrum shown in Fig.~\ref{fig:eresp-n}.
%
\section {Measurement of neutron spectrum in $^{12}$C~+~$^{93}$Nb reaction at
E($^{12}$C)~=~40~MeV}

\begin{figure}[!t]
\begin{center}
\includegraphics[scale=0.75]{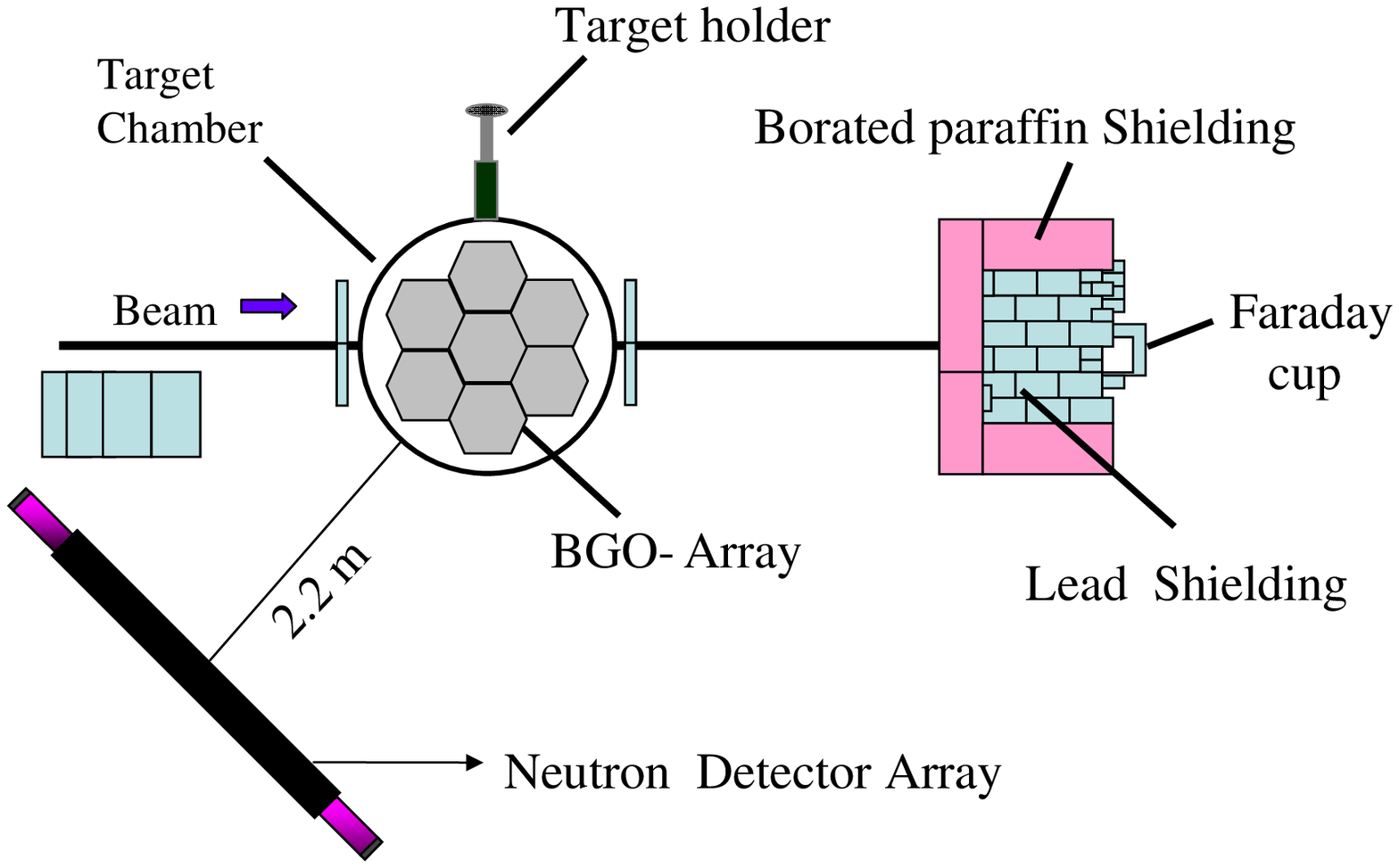}
\caption{A schematic of  the experimental setup used in the
$^{12}$C~+~$^{93}$Nb experiment.}
\label{fig:expsetup}
\end{center}
\end{figure}

A collimated, pulsed $^{12}$C beam (9.4 MHz) of  40 MeV from the
Mumbai Pelletron bombarded a self supporting $^{93}$Nb target of
thickness 0.5 mg/cm$^{2}$. Neutron spectra were measured with 15 of
the 16 plastic scintillators. The detector array was placed at 2.2~m
from the target and at 135$^\circ$ with respect to the beam
direction. It was covered with a 25~mm thick lead shield. The beam
collimators and the beam dump were shielded with lead bricks and
borated paraffin blocks to reduce the gamma and neutron background.
The measurements were made in coincidence with low energy
$\gamma$-rays originating from the yrast transitions of the fusion
residues. These   $\gamma$-rays were detected in  an array of 14
bismuth germanate (BGO) detectors~\cite{mitra1}. The BGOs were
shielded from background gamma rays coming from the collimators and
beam dump with suitably placed lead bricks. The schematic
experimental setup is shown in Fig.~\ref{fig:expsetup}.
%

\begin{figure}[!t]
\begin{center}
\includegraphics[scale=0.90]{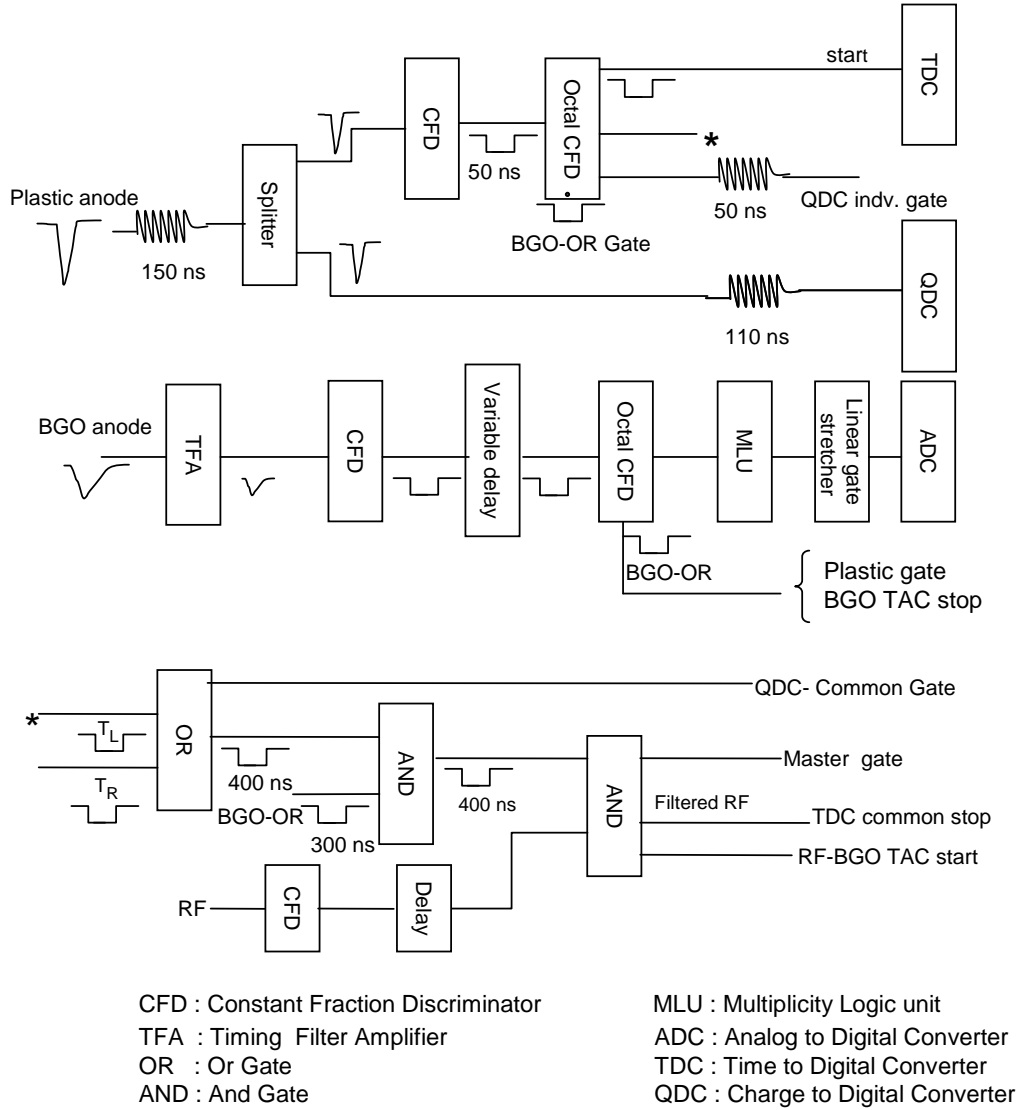}
\caption{Block diagram of the electronics set up used in the
$^{12}$C+$^{93}$Nb experiment.}
\label{fig:electronics}
\end{center}
\end{figure}

The total efficiency of BGO array for 662 keV $\gamma$-ray, measured
using a $^{137}$Cs source kept at the target position, was
$\sim$~65\%. The schematic electronic block diagram is shown in
Fig.~\ref{fig:electronics}. The high voltage applied to the PMTs of
individual plastic scintillator detectors was adjusted to provide
the same pulse height for 4.44 MeV $\gamma$-ray from $^{241}$Am-Be
source placed at the centre of the detector. The full energy range
was E$_{ee}$~$\sim$~28 MeV. The signal from the anode was split into
two parts. One was fed to a QDC for the energy measurement and the
other was sent to a constant fraction discriminator~(CFD) to
generate the timing signal. The CFD thresholds were  set at
$\sim$100 keV. One of the CFD outputs was sent to the corresponding
start channel of the time to digital converter (TDC) filtered
through the condition that at least one of the BGO detectors fire.
The second CFD output was used to generate the QDC gate. The third
output was used to define a valid event by combining signals from
all the scintillators in a logic OR unit. This OR
output~(Plastic-OR) was used for the common QDC gate.

\begin{figure}[!t]
\begin{center}
\includegraphics[scale=0.9]{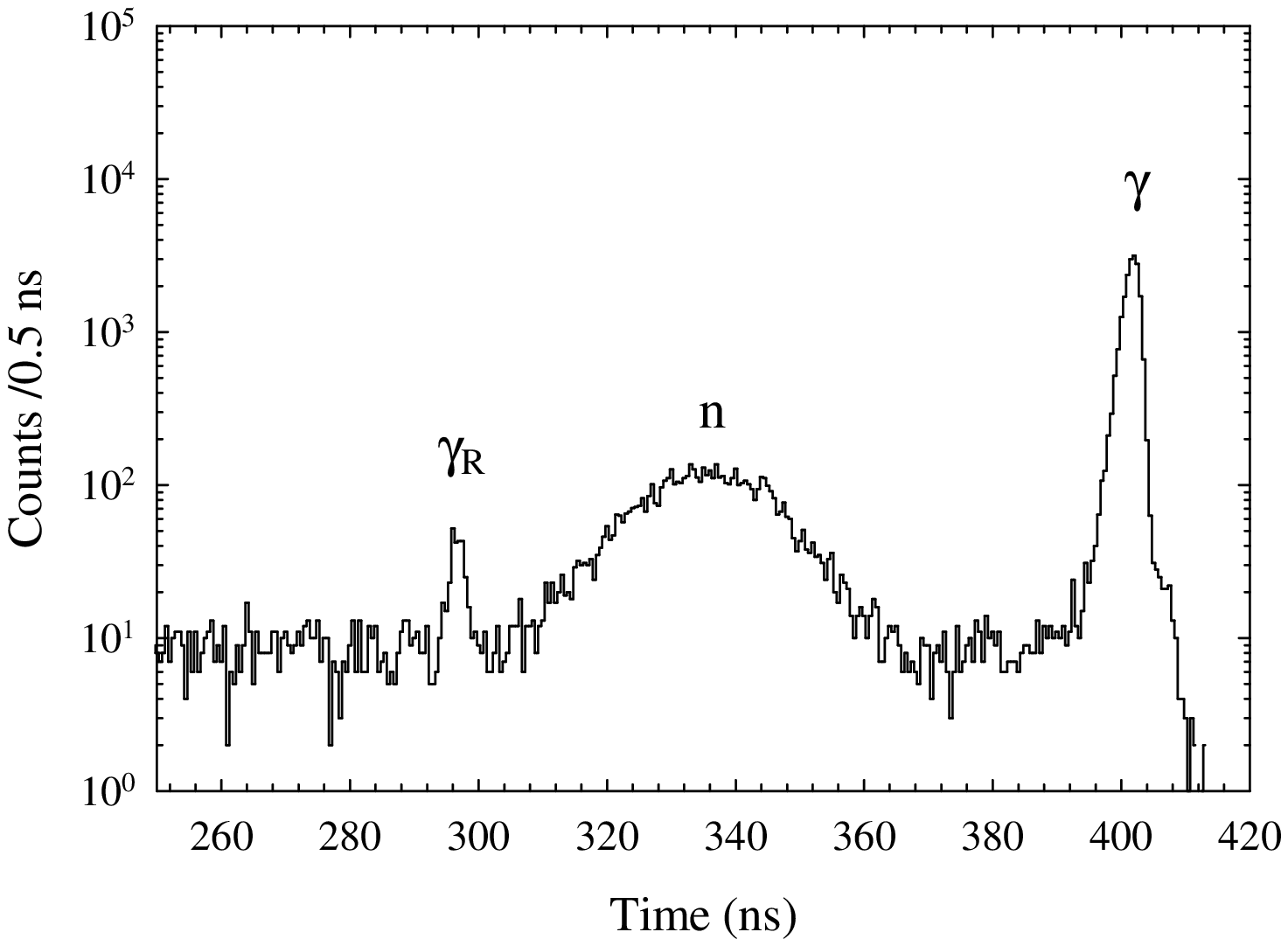}
\caption{Measured TOF spectra in the reaction $^{12}$C+$^{93}$Nb at
E($^{12}$C)=40~MeV for  BGO fold $\ge$1. The prompt gamma peak
($\gamma$), the broad bump due to neutrons (n) and the gamma peak
 from the adjacent beam burst ($\gamma _R$) can  be seen.}
\label{fig:cnb-tof-f1-14}
\end{center}
\end{figure}

\begin{figure}[!t]
\begin{center}
\includegraphics[scale=0.90]{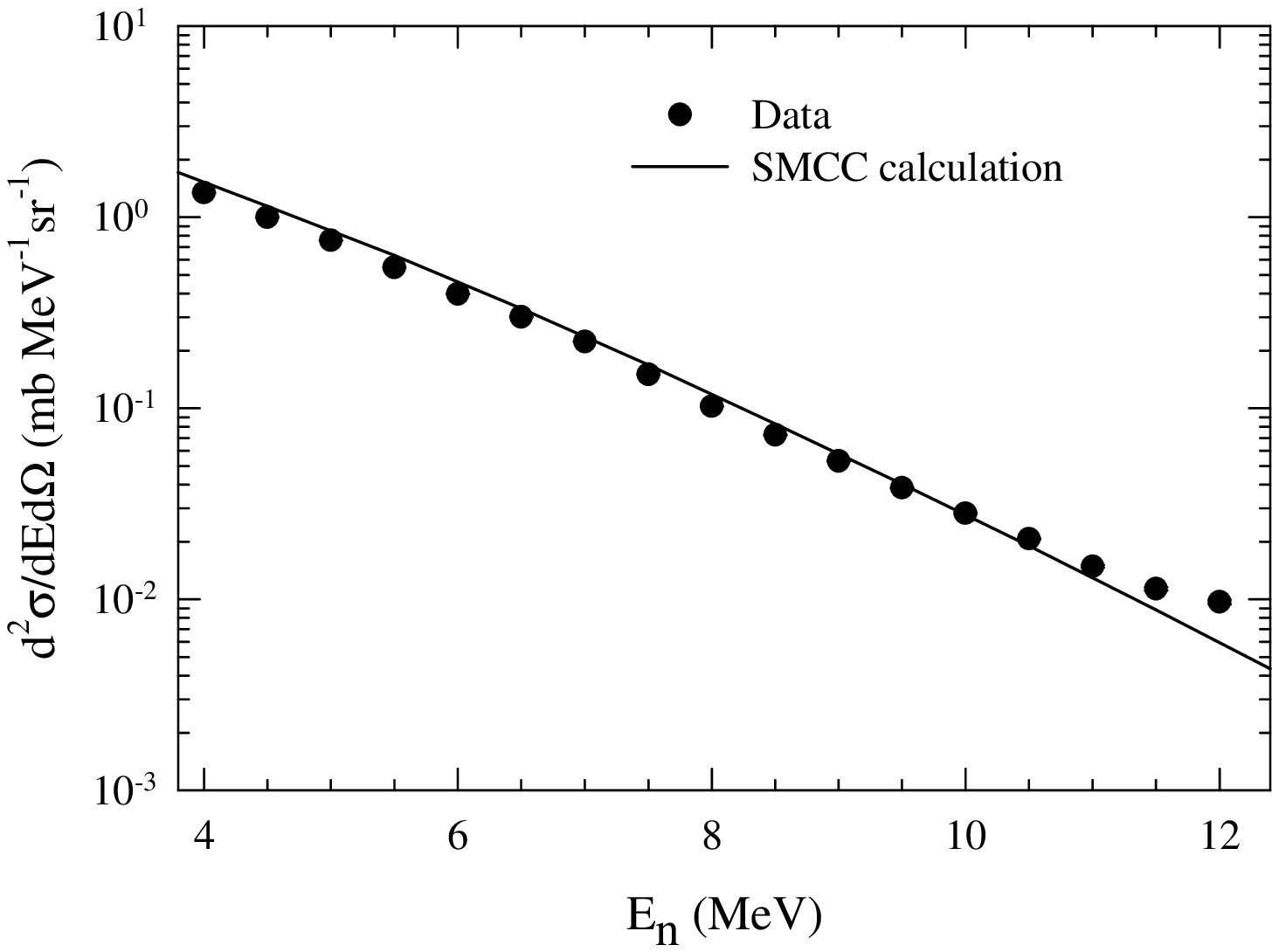}
\caption{Measured energy differential neutron cross sections in the
reaction $^{12}$C+$^{93}$Nb at E($^{12}$C)=40~MeV  for fold~$\ge$1
and statistical model calculations (SMCC) multiplied by 0.7.}
\label{fig:enspec-cnb}
\end{center}
\end{figure}

The timing signals from the BGO detectors were time matched and sent
to a multiplicity logic unit~(MLU) \cite{gothe}. The time matched
OR-output (BGO-OR) was used to filter the plastic CFD signals as
mentioned above. A coincidence between the BGO-OR and the Plastic-OR
was used to filter the RF signal derived from the beam pulsing
system. The filtered RF was used as the common stop for all the TDC
channels. The left and right times of each plastic scintillator were
measured with respect to the filtered RF. In addition, the
integrated charges (Q$_L$, Q$_R$), number of BGO detectors firing
simultaneously (fold) and the RF-BGO time were recorded in an event
by event mode. The time and energy calibrations were done as
mentioned earlier.
%
In the offline analysis, the TOF was derived from T$_L$ and T$_R$.
The zero of the time scale was determined from the prompt gamma peak
in the TOF spectrum. The position information was obtained from the
difference between the left and right time. Four position gates were
decided by inspecting the position spectra. A TOF spectrum for a
typical plastic scintillator with position gating on a quarter of
its length and at least one BGO in coincidence is shown in
Fig.~\ref{fig:cnb-tof-f1-14}. The random background reduces
drastically with the BGO coincidence requirement allowing the
measurement of neutrons with low production cross section. The fold
gated neutron time of flight spectra, after subtracting the random
background for each position cut, were converted to energy spectra.
%
The energy dependent efficiencies of the neutron detectors of the
array for different position gates were calculated using the Monte
Carlo simulation program mentioned earlier. The energy threshold
used in the simulation was the same as that used in deriving the
experimental spectra. The calculated efficiencies were used to
obtain the energy differential cross-sections in the c.m. system.
The c.m. spectrum for each fold was derived as the average of the
corresponding spectra from all the 15 detectors and four positions.
The energy differential cross sections in the c.m. system for
fold~$\ge$1 is shown in Fig.~\ref{fig:enspec-cnb}.
%

%
\begin{figure}[!t]
\begin{center}
\includegraphics[scale=0.90]{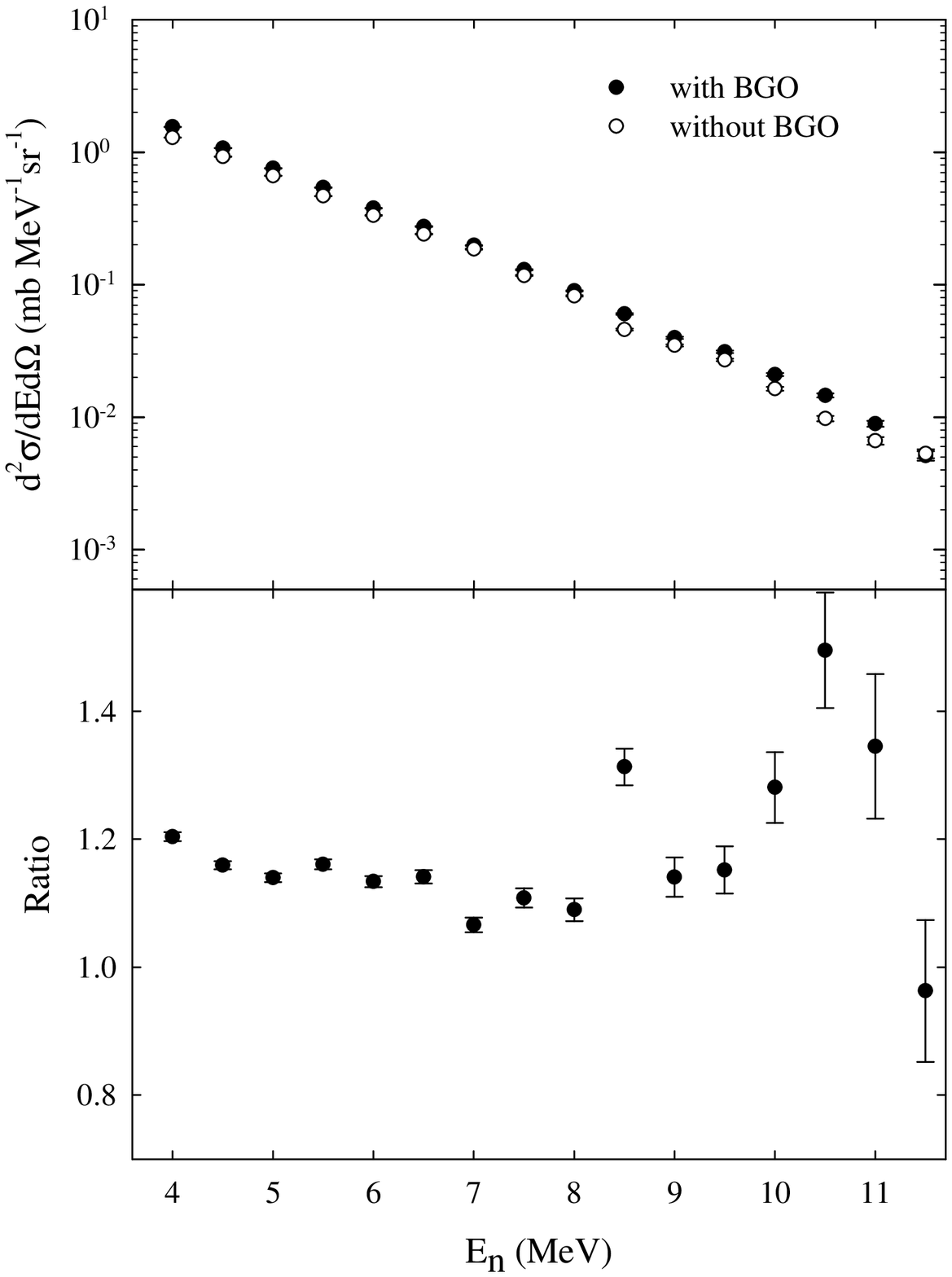}
\caption{Measured energy differential neutron cross sections in the
reaction $^{12}$C+$^{93}$Nb at E($^{12}$C)=40~MeV  for fold~$\ge$1
with and without top BGO array and the ratio between them at
different neutron energies.}
\label{fig:nscat}
\end{center}
\end{figure}
A statistical model calculation using the simulated Monte Carlo
CASCADE (SMCC)~\cite{drc} code was performed to compare with the
measured spectra. The calculation, done for the projectile energy
39.5 MeV (at the centre of the target) and a fusion cross-section of
188 mb, yielded the population matrix $ \sigma(E_n, J_{res})$ where
$ J_{res}$ refers to the residue spin. The $ J_{res}$ to F (fold)
response function was calculated using a Monte Carlo program with
BGO efficiency and inter-detector cross talks as inputs and assuming
the multiplicity$ M = J_{res}/1.9 + M_{0}$ with $M_{0}$=1.4. Using
this response function, the fold gated neutron spectra were obtained
from $ \sigma(E_n, J_{res})$. The shape of the experimental fold
gated neutron spectra agree well with those given by the SMCC
calculation for a level density parameter of A/8.5 MeV$^{-1}$. This
level density parameter is similar to that derived from the analysis
of the proton spectra from an earlier measurement in the same system
\cite{mitra1}. Fig.~\ref{fig:enspec-cnb} shows the calculation for
F$\ge$1 with an overall multiplication factor of 0.7.

The presence of the BGO detector array also contributes to scattered
neutrons some of which are detected in the plastic array. In order
to assess this contribution, the neutron TOF  measurements were
performed with and without the top BGO detector array. Only the
lower BGO array was used to generate the gamma ray multiplicity in
both the cases. The measured differential neutron cross sections and
their ratio for these two cases are shown in Fig~\ref{fig:nscat}.
The presence of the top BGO array gives rise to an additional
$\sim$13\% contribution, almost independent of neutron energy, which
is in reasonable agreement with the Monte Carlo simulation. This
implies an additional $\sim$26\% contribution due to the presence of
the full BGO $\gamma$-multiplicity array. Thus the measured cross
sections shown in Fig.~\ref{fig:enspec-cnb} get reduced by a factor
of $\sim$1.26 and become $\sim$1.8 times lower than the calculated
values. This discrepancy could arise from 1)~the uncertainty in
fusion cross section used in the SMCC calculation and 2)~the
presence of low energy electromagnetic transitions in the residues
which are detected with poor efficiency and, hence, affect the
J$_{res}$ to M prescription used in the calculation.
\section{Summary}
In summary, we have set up a 1$\times$1 m$^2$ plastic detector array
for neutron time of flight measurements at the BARC-TIFR Pelletron
laboratory. We have measured the response of the detector to
electrons using radioactive sources and mono-energetic neutrons
using the $^7$Li(p,n$_1$) reaction. A Monte Carlo simulation program
has been developed to simulate the response of the detector for
neutrons and is in agreement with our measurements. The detector
array was used to measure the energy differential neutron cross
sections in the reaction $^{12}$C~+~$^{93}$Nb at
E($^{12}$C)~=~40~MeV in coincidence with a 14-BGO gamma multiplicity
array. The measured spectral shape is in reasonable agreement with a
statistical model calculation.

\noindent {\bf Acknowledgements}\\
We thank the Pelletron staff for accelerator operation, Mechanical
Design and Prototype Development Section, BARC, for fabricating the
mechanical stand and the lead shield and H. H. Oza and M. S. Pose
for their help during the experiment. We would also like to thank
the anonymous referee for his critical comments.

\end{document}